\documentclass[%
 reprint,
 amsmath,amssymb,
 aps,
showkeys,
]{revtex4-2}

\usepackage{graphicx}
\usepackage{dcolumn}
\usepackage{bm}

\begin{document}

\preprint{APS/123-QED}

\title{DETERMINATION OF ASYMPTOTIC NORMALIZATION COEFFICIENTS \\ FOR THE $^{7}$Li$\to \alpha + ^{3}$He CHANNEL}

\author{L.D. Blokhintsev}
 \homepage{http://blokh@srd.sinp.msu.ru}
\affiliation{Skobeltsyn Institute of Nuclear Physics, Lomonosov Moscow State University\\Leninskie Gory, 1, bldg. 2,  Moscow, 119991, Russia
}%

\author{B.F. Irgaziev}
\homepage{http://irgaziev@yahoo.com}
\affiliation{Institute of Theoretical Physics, National University of Uzbekistan\\4 University Street, Tashkent,100174, Uzbekistan
}%

\author{D.A. Savin}
 \homepage{http://savin@srd.sinp.msu.ru}
\affiliation{Skobeltsyn Institute of Nuclear Physics, Lomonosov Moscow State University\\Leninskie Gory, 1, bldg. 2,  Moscow, 119991, Russia
}%

\date{\today}

\begin{abstract}
Asymptotic normalization coefficients (ANCs) $C_{3/2}$ and  $C_{1/2}$ for the channels $^7$Li$(3/2^-;0$ MeV)$\to \alpha+^3$H  and $^7$Li$(1/2^-;0.478$ MeV)$\to \alpha+^3$H, respectively, were determined by analyzing elastic $\alpha^3$H- scattering data using three different methods. All three methods yield similar results. The ANC values averaged over the three methods are $C_{3/2}=2.08\pm 0.10$ fm$^{-1/2}$ and $C_{1/2}=2.00\pm 0.10$ fm$^{-1/2}$. Comparison of the found ANCs with the previously obtained ANCs for $^7$Be confirms the relationship linking the ANC values for mirror nuclei. This research on determining the ANCs for the $^7$Li  nucleus is, to some extent, a continuation of the previously completed study on its mirror nucleus, $^7$Be   [ D. A. Savin, L. D. Blokhintsev, B. F. Irgaziev, A. S. Kadyrov, and A. M. Mukhamedzhanov, Phys. Rev. C 112, 065807 (2025)].
\end{abstract}


\maketitle

\section{Introduction}

Asymptotic normalization coefficients (ANCs) that determine the asymptotic behavior of wave functions in binary channels are widely used in modern nuclear physics, especially when analyzing data on nuclear reactions. ANCs are on-shell observables in contrast to standard spectroscopic factors. The definition of ANCs, their properties and methods for their determination from experimental data can be found in the review papers \cite{MB,Mukh}. It should be emphasized that theoretical ANC values are highly sensitive to the nuclear models used for their calculation. Different models that describe nuclear characteristics such as binding energies and root-mean-square radii equally well can yield significantly different ANC values (see, for example, \cite{Zhu}).
ANCs are especially important in determining the cross sections for astrophysical nuclear reactions, which are inaccessible for direct measurement due to the large Coulomb barrier. In particular, ANCs determine the astrophysical factors for radiative capture reactions at astrophysical energies \cite{Mukh1,Xu,Mukh2,Mukh3}.

Radiative capture reactions $^3$H$(\alpha,\gamma)^7$Li and $^3$He$(\alpha,\gamma)^7$Be play an important role in nuclear astrophysics. These processes lead to the formation of the primordial $^7$Li in the early Universe \cite{Burles,Nollett1,Nollett2}. Information about their cross sections is important, in  particular, due to the existence of the so called cosmological lithium puzzle. This puzzle lies in that the theoretical calculations of the Big Bang nucleosynthesis processes, which match the observed primordial abundances of hydrogen and helium isotopes with remarcable precision, lead to the calculated quantities for $^7$Li which are 3 to 4 times higher than the observed values (see e.g. \cite{lithium} and references therein). Reaction $^3$He$(\alpha,\gamma)^7$Be, in addition, is a primary reaction for chains II and III of the $pp$-cycle of hydrogen burning in stars \cite{Cauldrons}. 

The energy range for the primordial nucleosynthesis of light isotopes such as $^2$H, $^{3,4}$He, and $^7$Li  lies between 30 to 100 keV. In this context, the ANCs play a crucial role in driving the reactions involved in this primordial nucleosynthesis.

The $^3$He$(\alpha,\gamma)^7$Be reaction has been studied in a number of experimental and theoretical works (see \cite{Adel,Wiescher,SBIKM} and references therein). From these works it follows that the cross section of this reaction is largely determined by the values of the ANCs for the channels $^7$Be$(3/2^-;0$ MeV)$\to \alpha+^3$He  and $^7$Be$(1/2^-;0.429$ MeV)$\to \alpha+^3$He. The values of these ANCs are discussed in detail in \cite{SBIKM}. The mirror reaction $^3$H$(\alpha,\gamma)^7$Li has been studied much less, and information about the ANCs corresponding to the virtual decay of $^7$Li into $^3$H and $\alpha$ is quite meager.

The present paper is devoted to the determination of the ANCs for the
 $^7$Li nucleus in the $\alpha+^3$H channel by different methods from the analysis of elastic $\alpha^3$H-scattering data. The  $^7$Li  nucleus has two bound states: the ground state with $J^\pi=3/2^-$  and the excited state with $J^\pi=1/2^-$   and the excitation energy 0.478 MeV. The binding energies of these states in the $\alpha+^3$H channel are $\varepsilon_{3/2}=2.467$ MeV and $\varepsilon_{1/2}=1.989$ MeV. The ANCs for these states will be designated as $C_{3/2}$ and  $C_{1/2}$, respectively. As the nuclei $^7$Li and $^7$Be are mirror images of each other, it is important to compare their ANCs using the method outlined in the paper \cite{MB}. The present paper  can be seen as a natural continuation of \cite{SBIKM}.

The remainder of this paper is organized as follows. Section II contains formulas necessary for understanding the further text. The results for the ANCs obtained by different methods are presented in Section III. These results are briefly discussed in Section IV.

We use the system of units in which $\hbar = c = 1$.

\section{Basic formalism}

 In this section, we present some formulas related to the analytical continuation to the region of negative energies of the elastic nuclear scattering amplitude in the presence of Coulomb interaction. A more detailed presentation can be found, for example, in \cite{BKMS5,BKMS6}. 

The partial component of the Coulomb-renormalized scattering amplitude has the form \cite{Hamilton}
\begin{equation}\label{renorm}
\tilde f_l(E)=\frac{\exp(2i\delta_l)-1}{2ik}C_l^{-2}(\eta) \equiv \frac{k^{2l}}{D_l(E)},
\end{equation}
where $\delta_l$ is the Coulomb-nuclear phase shift, $k=\sqrt{2\mu E}$ is the relative momentum, $l$ is the orbital angular momentum,    $\eta =Z_bZ_ce^2\mu/k$ is the Coulomb  parameter for the $bc$ scattering state, $\mu$ is the reduced mass, $Z_ie$  is the electric charge of particle $i$, and $C_l(\eta)$ is the  Gamow factor:   
\begin{align}\label{C}
C_l(\eta)&=\left[\frac{2\pi\eta}{\exp(2\pi\eta)-1}v_l(\eta)\right]^{1/2}, \\  
v_l(\eta)&=\prod_{n=1}^{l}(1+\eta^2/n^2)\;(l>0),\quad v_0(\eta)=1.
\end{align}
The function $D_l(E)$ in Eq. \eqref{renorm} can be written as
\begin{equation}\label{D}
D_l(E)=\tilde \Delta_l(E)-ik^{2l+1}C_l^2(\eta)
\end{equation}
or, equivalently, as
\begin{equation}\label{K}
D_l(E)=K_l(E)-2\eta k^{2l+1} h(\eta) v_l(\eta),
\end{equation}
where $K_l(E)$ is the Coulomb-modified effective-range function (ERF) 
\begin{equation}\label{ERF}
K_l(E)= k^{2l+1} \left[ C_l^2(\eta)(\cot\delta_l-i) + 2 \eta h(\eta)v_l(\eta) \right],
\end{equation}
\begin{equation} \label{h}
h(\eta) = \psi(i\eta) + \frac{1}{2i\eta}-\ln(i\eta),   
\end{equation}
$\psi(x)$ is the digamma function and
\begin{equation} \label{Deltal}
\tilde\Delta_l(E)=k^{2l+1}C_l^2(\eta)\cot\delta_l.
\end{equation}

 ${\tilde f}_{l}(E)$ has no essential singularity on the physical sheet of energy and can be analytically continued into the  region  $E<0$ \cite{Hamilton}.

We are interested in the case when the $b+c$ system has the bound state $a$ with the binding energy $\varepsilon=\kappa^2/2\mu>0$ in the partial wave $l$. In this case,  $\tilde f_l$ has a pole at $E=-\varepsilon$ and  the ANC $C_l(a\to b+c)$ is expressed through the residue of $\tilde f_l$ at this pole  \cite{BMS} 
\begin{align} \label{C1}
C_l^2&=-2\mu\left[\frac{\Gamma(l+1+\eta_b)}{l!}\right]^2
\mathrm{res} \, \tilde f_l(E)|_{E=-\varepsilon},\\
&=    
-2\mu\left[\frac{\Gamma(l+1+\eta_b)}{l!}\right]^2(-1)^l\varkappa^{2l}\left(\frac{dD_l(E)}{dE}\right)^{-1}_{E=-\varepsilon}.
\end{align}
where $\eta_b=Z_bZ_ce^2\mu/\kappa$ is the Coulomb  parameter for the bound state $a$ and $\Gamma(z)$ is the Gamma function.

To determine the ANC, you can use the ERF $K_l(E)$. $K_l(E)$ has no singularities at $E = 0$. It can be  analytically continued to $E=-\varepsilon$ and  the corresponding ANC can be determined by using Eqs. (5) and (10) (ERF method). However, the disadvantage of this method is that in practice it can only be applied to light nuclear systems due to the presence of a  background of purely Coulomb terms (see \eqref{K}, \eqref{h}). These terms increase in absolute value with increasing charge and mass of the nucleus, which significantly reduces the accuracy of the obtained ANC values.

In work \cite{Sparen}, to determine the ANCs, it was proposed to continue analytically the function $\tilde\Delta_l(E)$ \eqref{Deltal} instead of the ERF $K_l(E)$ ($\Delta$-method).  However, unlike the ERF method, the $\Delta$-method is not formally correct.            $\tilde\Delta_l(E)$ possesses the essential singularity at $E=0$ and cannot be represented by a convergent series in $E$ in the neighborhood of $E=0$. In addition, in the $\Delta$-method, the  condition $\tilde\Delta_l(E)=0$ is used as the amplitude pole condition, which does not coincide with the exact pole condition $D_l(E)=0$. However, despite its lack of rigor, this method can be used with acceptable accuracy for nuclear systems with large charge and mass. It was shown in Refs. \cite{BKMS2,Gaspard} that the $\Delta$-method can be employed to obtain reliable information on a bound state if the corresponding binding energy and the  energy of scattering states used to approximate the $\tilde\Delta$ function satisfy the condition
  \begin{equation}\label{range}
|E|\le (Z_bZ_ce^2)^2\mu/2=1\,\mathrm{Ry},
\end{equation}
where 1 Ry is the nuclear Rydberg energy. 

A new method of analytical continuation called the subtraction method (abbreviated as S-method) was proposed in \cite{BKMS6}. This method, unlike the two methods described above, can be applied to nuclear systems of arbitrary mass. It was first used in Ref. \cite{SBIKM} to determine the ANCs for the $^7$Be nucleus from $^4$He$+^3$He scattering data. The method is based on the analytical continuation of the function $g_l(E)$, which can be written in the following equivalent forms
\begin{align} \label{g}
g_l(E)&= D_l(E)-D_l^{SW}(E)\\
&=\tilde\Delta_l(E)-\tilde\Delta_l^{SW}(E)\\
&=K_l(E)-K_l^{SW}(E),
\end{align}
where $D_l^{SW}(E)$, $\tilde\Delta_l^{SW}(E)$ and $K_l^{SW}(E)$ are the functions $D_l(E)$, $\tilde\Delta_l(E)$ and $K_l(E)$ for a potential that is the sum of the Coulomb and square well (SW) potentials. The functions $\tilde\Delta_l^{SW}(E)$ and $K_l^{SW}(E)$ are  expressed explicitly in terms of parameters of the SW potential (see Eq. (16) from Ref. \cite{BKMS5}). 
$K_l(E)$ and $K_l^{SW}(E)$ have no essential singularity at $E=0$. Hence $g_l(E)$ can be approximated by a polynomial in $E$ at $E>0$  and continued analytically into the negative-energy region. This procedure can be implemented with arbitrary parameters of the SW potential. Note  that $g_l(E)$ is real for both positive and negative energies and does not contain purely Coulomb terms. 

The condition of the pole of $\tilde f_l(E)$ at $E=-\varepsilon=-\varkappa^2/2\mu$ is as follows
\begin{equation}\label{cond}
D_l(-\varepsilon)=g_l(-\varepsilon)+D_l^{SW}(-\varepsilon)=0. 
\end{equation}
ANC $C_l$ is determined by the expression (10).

The results of determining the ANCs for $^7$Li obtained by the S-method are presented in the next section along with the results obtained by other methods.

\section{Determination of the ANCs $C_{3/2}$ and $C_{1/2}$}
	
All methods for determining the ANCs used in this work are based on information about the phase shifts of elastic $\alpha^3$H scattering  in the  channels $3/2^-$ and $1/2^-$. In the published sources there are only two considerably old sets of low-energy $\alpha^3$H  scattering phase shifts: i) M. Ivanovich et al. \cite{Ivan} (set A) and ii) R. J. Spiger et al. \cite{Spiger} (set B). The phase-shift values from these sets differ significantly from each other for both channels $3/2^-$ and $1/2^-$.  This difference is demonstrated in Fig. 1 for the $3/2^-$ channel. The solid line in Fig. 1 corresponds to the results of phase shift calculations within the framework of a potential model in which the $\alpha^3$H interaction is described by a potential in the form of the sum of the nuclear potential of a square well (SW) and the Coulomb potential. The parameters of the SW potential are the same as those found in  \cite{SBIKM} from the requirement of the best description of the phase shifts of the $\alpha^3$He scattering process. From Fig. 1 it is evident that this potential describes the  $\alpha^3$H scattering phase shifts of set A quite well. This is a natural result, since systems  $\alpha^3$He  and $\alpha^3$H are mirror systems. On the other hand, this potential's description of the phase shifts of set B is clearly unsatisfactory. For the $1/2^-$ channel the situation is similar. Taking into account the above, we will rely specifically on the phase shifts of set A \cite{Ivan} from now on. 

\begin{figure}[htb]
\includegraphics[scale=1.0]{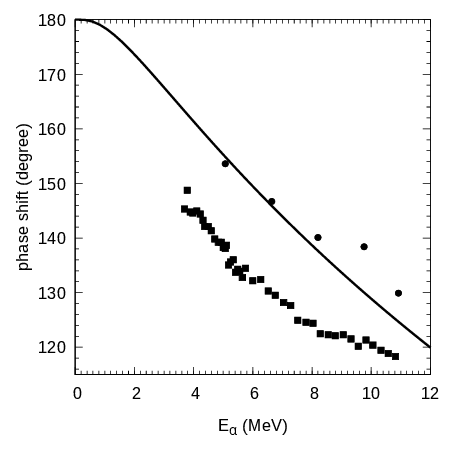} 
\caption {$\alpha^3$H-scattering phase shifts for the $3/2^-$ state. Circles - \cite{Ivan}, squares - \cite{Spiger}. Solid line - phase shifts calculated with the potential for $^7$Be.}
\label{fig1}
\end{figure}

Set A includes 5 phase shifts in the range of incident $\alpha$-particle energy $E_\alpha=5.07-10.92$ MeV, which corresponds to the range of c.m. energy  $E=2.19-4.69$ MeV. The values of these phase shifts are given in \cite{Ivan} without indicating the experimental errors. To better reproduce the real experimental situation, we superimposed these phase shifts with a random error of
5\% as in \cite{BKMS5}. 

Below in this section we present the results for the ANCs $C_{3/2}$ and  $C_{1/2}$ obtained by different methods based on indicated phase shifts. We do not use the $\Delta$-method, since for the $\alpha+^3$H system  1 Ry = 0.17 MeV and the used values of scattering energy and binding energies sharply violate the condition \eqref{range} of applicability of this method. As for the ERF-method, the results obtained using it are unstable with increasing degree of the polynomial approximating the ERF. This instability may be due to the small number of phase shifts presented in \cite{Ivan}.

\subsection{$R$-matrix approach}
The single-level one-channel $R$-matrix is of the form
\begin{equation} \label{R}
R(E)= \dfrac{\gamma^2}{E_R-E}.
\end{equation}
If in a given channel the system has a single bound state with binding energy $\varepsilon$, then in order for the scattering amplitude to have a pole at $E = -\varepsilon$, it is necessary to put $E_R = -\varepsilon$. For the $\alpha+^3$H system, $\varepsilon=2.467$ MeV and $\varepsilon=1.989$ MeV for channels $3/2^-$ and $1/2^-$, respectively.

 The phase shift $\delta(E)$ is expressed through the reduced width $\gamma^2$ and the channel radius $a$ \cite{Rmatrix1}. Applying the criterion $\chi^2$ to Eq. \eqref{R}  for a fixed $a$ results in the relation
\begin{equation} \label{R1}
\gamma^2 \sum_{i=1}^n  \dfrac{1}{\epsilon_i^2 (E_R-E_i)^2}   = \sum_{i=1}^n \dfrac{R_i}{\epsilon_i^2(E_R-E_i)},  
\end{equation}
where $E_i$ is the value of   energy $E$ for which the phase shift $\delta_i$ is known, $R_i = R(E_i)$, 
$\epsilon_i$ is the error of $R_i$ determined by the error of $\delta_i$.  Eq. \eqref{R1} allows us to express the reduced 
width $\gamma^2$ in terms of the other quantities included in \eqref{R1}, which depend on $a$. Then using the criterion $\chi^2$  with respect to $a$, it is possible to determine the parameters $a$ and $\gamma^2$ that lead to the best description of the results of the phase-shift analysis. 

Applying the above procedure to the phase shifts of set A \cite{Ivan} results in the following $R$-matrix parameters
\begin{align} \label{par}
& J^\pi=\frac{3}{2}^-: a=2.807\;\mathrm{fm},\; \gamma^2=3.718\;\mathrm{MeV}; \nonumber \\
& J^\pi=\dfrac{1}{2}^-: a=2.837\;\mathrm{fm},\; \gamma^2=6.570\;\mathrm{MeV}.  
\end{align}
$R$-matrices with parameters \eqref{par} describe the phase shifts of the set A quite well for both channels (see red lines in Figs. 2 and 3).
For the one-level $R$-matrix \eqref{R} with $E_R = -\varepsilon$ the ANC can be directly expressed through $\gamma^2$ and $a$ \cite{Rmatrix1,Mukh3,BS}. The $R$-matrix parameters given in \eqref{par} lead to $C_{3/2}=2.02$ fm$^{-1/2}$ and $C_{1/2}=2.01$ fm$^{-1/2}$.

\begin{figure}[htb]
\includegraphics[scale=1.0]{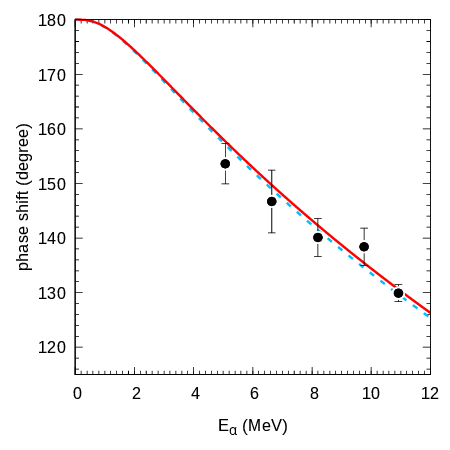} 
\caption {$3/2^-$ phase shifts for the optimal $R$-matrix
(solid red line) and for the SW potential (dashed blue line).}
\label{fig2}
\end{figure}

\begin{figure}[htb]
\includegraphics[scale=1.0]{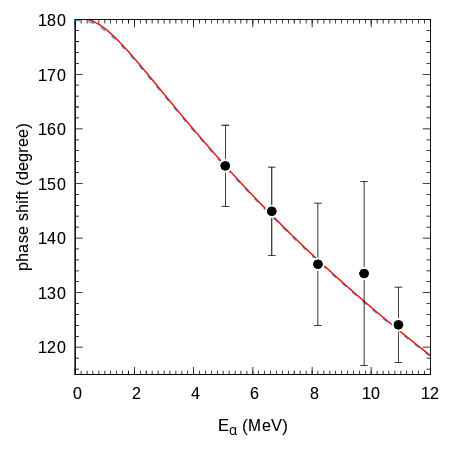} 
\caption {$1/2^-$ phase shifts for the optimal R-matrix
(solid red line) and for the SW potential (dashed blue line)}
\label{fig3}
\end{figure}

\subsection{Two-body potential model}
In this subsection, the ANCs $C_{3/2}$ and $C_{1/2}$ are found by solving the two-body Schr\"odinger equation with the potential in the form of the sum of the Coulomb potential and the square-well (SW) potential. In accordance with the ideas of the shell model,  the chosen potential leads to two bound states, the lower of which is forbidden. The radius $R$ and depth $V_0$ of the SW potential are found by the $\chi^2$ method from the condition of the best description of the experimental phase shifts along with the correct reproduction of the  binding energies of the ground ($3/2^-$) and excited ($1/2^-$) states of $^7$Li.  The above coinditions result in the following parameters  
\begin{align} \label{SW}
& J^\pi=\frac{3}{2}^-: R=3.94\;\mathrm{fm},\; V_0=58.89\;\mathrm{MeV}; \nonumber \\
& J^\pi=\dfrac{1}{2}^-: R=3.28\;\mathrm{fm},\; 
V_0=50.31\;\mathrm{MeV}.  
\end{align}
Parameters \eqref{SW}  lead to $C_{3/2}=2.10$ fm$^{-1/2}$ and $C_{1/2}=1.98$ fm$^{-1/2}$.

It should be noted that the SW potentials with parameters \eqref{SW}, despite their simplicity, describe the experimental data quite well (see blue lines in Figs. 2  and 3). 

\subsection{S-method}
The basics of the recently proposed S-method are outlined above in Section 2. Within  this method, function $g_l(E)$ in the form (13) is approximated at $E>0$ by polynomials in $E$ using the $\chi^2$ method and analytically continued to the pole point of the amplitude $\tilde f_l(E)$; the ANC values are then found using Eqs. (4) and (10). As noted in Section 2, the S-method is formally correct for any parameters of the SW potential that determines the function  $\tilde\Delta_l^{SW}(E)$. We chose the same parameters that, within the  two-body potential model, lead to the best description of the experimental data (see subsection 3.2). 
The procedure described results in $C_{3/2}=2.11$ fm$^{-1/2}$ and $C_{1/2}=2.01$ 
fm$^{-1/2}$. 

Note that the results for $C_{3/2}$ and $C_{1/2}$ obtained with direct use of the SW potentials (see subsection 3.2) and within the S-method using the SW potentials with the same parameters are practically the same. This is due to the fact that the SW potential describes the experimental data very well. As a result, the difference function $g_l(E)$ turns out to be small.

\subsection{Comparing ANCs for mirror nuclei $^7$Li and $^7$Be}
Let us introduce the Coulomb-renormalized ANC $\tilde C_i$, related to the standard ANC $C_i$ as 
\begin{equation} \label{renorm1}
C_i = R_{li} \tilde{C}_i,
\end{equation} 
where 
\begin{equation} \label{R2}
R_{li}=\dfrac{\Gamma(l+1+\eta_i)}{\Gamma(l+1)},
\end{equation}
$i=1$ and $i=2$ denote nuclei $^7$Li and $^7$Be, respectively, $\eta_i$ is the corresponding Coulomb factor. 
Generalizing Eq. (130) from the review \cite{MB} to the case where the Coulomb interaction is present in both channels, we can obtain the following approximate relation that the ANCs for mirror nuclei $^7$Li and $^7$Be should satisfy: 
\begin{equation} \label{ratio}
\dfrac{\tilde C_2}{\tilde C_1} \approx \dfrac{\kappa_1}{\kappa_2} \dfrac{R_{l1}}{R_{l2}} \dfrac{\varphi_2(R_{ch})}{\varphi_1(R_{ch})},
\end{equation}
where $\kappa_i$ is the wave number of the corresponding bound state, $\varphi_i$ is the normalized radial bound-state wave function ($\int_0^{\infty} \varphi_i^2(r) dr =1$), and  $R_{ch}$ is the channel radius.

Let us check the validity of Eq. \eqref{ratio} by comparing the ANC values for $^7$Li and $^7$Be, obtained respectively in the present work and in \cite{SBIKM} within the potential model. Let's start with state $3/2^-$. The functions $\varphi_i(r)$ are calculated by solving the Schr\"odinger equation for the sum of the SW and Coulomb potentials. The SW-potential parameters for $^7$Li are given above in subsection 3.2. For $^7$Be, the corresponding parameters are taken from \cite{SBIKM}. Functions $\varphi_1(r)$ and $\varphi_2(r)$ turn out to be very close to each other. Their overlap integral is equal to 0.975. Using the value $C_{1;3/2}=2.10$ fm$^{-1/2}$ obtained in the present  work and the value $C_{2;3/2}=2.73$ fm$^{-1/2}$ from \cite{SBIKM}, we obtain that the left-hand side of \eqref{ratio} is equal to 0.996. In Fig. 4 this value (blue horizontal line) is compared with the function on the right-hand side of \eqref{ratio} (black line). The red and green vertical lines indicate the radii of the SW potential for $^7$Li and $^7$Be, respectively. 

In Fig. 5, a similar comparison is made for state $1/2^-$, for which $\tilde C_2/\tilde C_1 = 0.934$. From Figs. 4 and 5 it follows that Eq. \eqref{ratio} is fulfilled with good accuracy if the channel radius is taken to be a value close to the radius of the nuclear potential.

\begin{figure}[htb]
\includegraphics[scale=1.0]{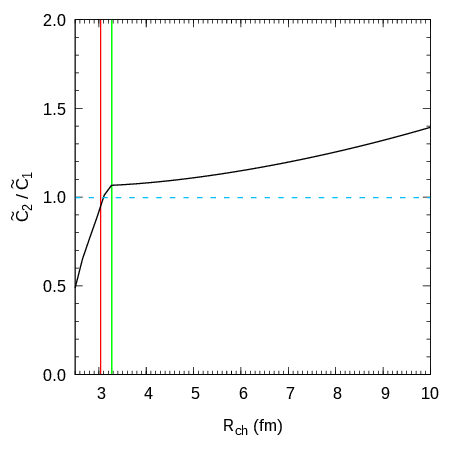} 
\caption {Checking Eq. \eqref{ratio} for the $3/2^-$ state. The meaning of the lines in the figure is explained in the text.}
\label{fig4}
\end{figure}

\begin{figure}[htb]
\includegraphics[scale=1.0]{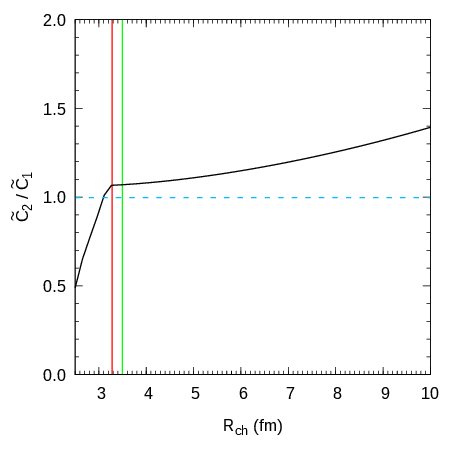} 
\caption {The same as in Fig. 4, but for the state $1/2^-$.}
\label{fig5}
\end{figure}

\section{Summary and conclusion}
In the present paper, the ANCs for the $^7$Li$\to\alpha+^3$H channel were determined using various methods. Both ground ($3/2^-$) and excited   ($1/2^-$) states of $^7$Li were considered. All methods used were based on data on phase shifts of elastic $\alpha^3$H scattering. For ANCs $C_{3/2}$ and $C_{1/2}$ the following results were obtained: $C_{3/2}=2.02$ fm$^{-1/2}$, $C_{1/2}=2.01$ fm$^{-1/2}$ 
($R$-matrix approach); $C_{3/2}=2.10$ fm$^{-1/2}$, $C_{1/2}=1.98$ fm$^{-1/2}$ (two-body potential model); $C_{3/2}=2.11$ fm$^{-1/2}$, $C_{1/2}=2.01$ fm$^{-1/2}$ (S-method). As can be seen, the results of different methods are very close. It is not possible to indicate sufficiently substantiated errors of the methods used.  However, if we assume that the errors in the obtained ANC values are the same 5\% that were imposed on the phase shifts used, then the ANC values averaged over the three methods are $C_{3/2}=2.08\pm 0.10$ fm$^{-1/2}$ and $C_{1/2}=2.00\pm 0.10$ fm$^{-1/2}$. A comparison of our ANCs for $^7$Li with the ANCs for   $^7$Be obtained by similar methods in \cite{SBIKM} shows that the approximate relation \eqref{ratio}, which connects the ANCs for mirror nuclei, is fulfilled with good accuracy. 

Let us compare the values of the ANCs obtained by us for $^7$Li with the results of other works. $C_{3/2}$ and $C_{1/2}$ were calculated in several theoretical works. In \cite{Vorabbi} and \cite{Rodkin}, within the framework of the no-core shell model, it was obtained
$C_{3/2}=3.49$ fm$^{-1/2}$, $C_{1/2}=3.16$ fm$^{-1/2}$ \cite{Vorabbi} and $C_{3/2}=3.44$ fm$^{-1/2}$, $C_{1/2}=2.95$ fm$^{-1/2}$ \cite{Rodkin}. To the best of the authors' knowledge, there is only one paper \cite{Igamov} in which the ANCs for $^7$Li were determined from the analysis of experimental data. Specifically, in this work, $C_{3/2}$ and $C_{1/2}$  were found by analyzing data on the astrophysical factor of the $^3$H$(\alpha,\gamma)^7$Li reaction at  energy $E\le 1200$ keV \cite{Brune} within the modified DWBA. As a result of the analysis, the following values were obtained: $C_{3/2}=3.57\pm 0.15$ fm$^{-1/2}$ and $C_{1/2}=3.00\pm 0.15$ fm$^{-1/2}$.

It is evident that the ANC values obtained in this work are somewhat lower than the values known in the literature. Comparison with the results of \cite{Igamov} confirms the conclusion made in \cite{SBIKM} for the $^7$Be$\to \alpha+^3$He channel that the ANCs extracted from elastic scattering data are consistently smaller than those obtained from radiative  capture measurements.
It should be noted that smaller values of the ANCs for $^7$Li and $^7$Be lead to smaller values of the cross sections of the reactions $^3$H$(\alpha,\gamma)^7$Li and $^3$He$(\alpha,\gamma)^7$Be  at astrophysical energies. 
 
 The decrease in these cross sections, in turn, means a decrease in the content of $^7$Li in the Universe which softens the cosmological lithium puzzle. To obtain more accurate information about the ANC values for $^7$Li, it is necessary to conduct detailed low-energy $\alpha^3$H-scattering measurements with careful phase-shift analysis using modern accelerators and computers.

\section{Acknowledgements} 
L.D.B. and D.A.S. note that the study was conducted under the state assignment of Lomonosov Moscow State University.

\end{document}